\documentclass[18pt,preprint]{aastex}
\begin{document}
\title{Spectroscopic observations of a sample of dwarf spiral galaxies. II-  Abundance gradients}
\author{A.M. Hidalgo-G\'amez}
\affil{Departamento de F\'{\i}sica, Escuela Superior de F\'{\i}sica y Matem\'aticas, IPN, U.P. Adolfo L\'opez Mateos, 
C.P. 07738, Mexico city, Mexico}
\author{D. Ram\'{\i}rez-Fuentes}
\affil{Instituto de Astronom\'\i a, UNAM, Ciudad Universitaria, Aptdo. 70 264,
C.P. 04510, Mexico City, Mexico}
\author{and J. J. G\'onzalez}
\affil{Instituto de Astronom\'\i a, UNAM, Ciudad Universitaria, Aptdo. 70 264,
C.P. 04510, Mexico City, Mexico}

\begin{abstract}

\keywords{ISM: abundances; (ISM): H\,{\sc ii} regions}

\end{abstract}

The oxygen gradient of four dS galaxies has been determined using abundances for 
several H\,{\sc ii} regions determined with four different methods. The gradient 
slopes of the three non-barred galaxies in the sample are quite steep, larger 
than $-0.2$ dex/kpc, while the gradient of the barred galaxy is shallower, only 
$-0.1$ dex/kpc. Although these gradients are quite steep they are real, following 
all the galaxies the same trend. Moreover, the results obtained here agree with those 
marked by the late-type, non-dwarf spirals, particularly the relationship between 
the gradient and the absolute magnitude and the optical size for non-barred 
galaxies, and the surface density for barred ones.

\section{Introduction}

The existence of differences in the chemical abundances along the galactocentric 
 distance is a well known characteristic of spiral galaxies (see, among others, the  
 works of Searle 1971 as well as  D\'{\i}az 1989; Zaritsky et al. 1989; Walsh \& Roy 
  1989; Vila-Costas \& Edmunds 1992). Normally, the central regions exhibit higher 
   abundances than those located at the outskirts. Such gradients, both nebular and 
    stellar, have been obtained for our Galaxy (e.g. Edvardsson 2001 and references 
     therein), but not without controversy (Fu et al. 2009). 

The gradients in metallicity are probably the results of billions of years of evolution. 
 As the chemical elements are processed in the interior of the stars, such abundance 
  gradients can be related to the gas mass fraction and the star formation rate in the 
   disc of spiral galaxies (Phillips \& Edmunds 1991) or to variations in the Initial 
    Mass Function  (G\"usten \& Mezger 1982). They also could be related with the yield 
     if the closed-box model is considered. A constant yield might give variations in 
      the gas fraction, and vice versa (J. V\'{\i}lchez, 2010 private communication). 

Several investigations studied the dependence of the gradient with different properties 
 of the galaxies, such as the morphological type, the absolute magnitude and the 
  rotation velocity (e.g. Zaritsky et al. 1994; Vila-Costas \& Edmunds 1992). The main 
   conclusion of these investigations is that gradients in non-barred galaxies seem to 
    be related to the mass surface density and to the morphological type (late-type 
     galaxies have steeper gradients than early-types) while the central abundance are 
      correlated with the galaxy mass. For barred galaxies  the gradients seem to be 
       related to the length of the bar compared to the size of the disc and to the 
        ellipticity (Martin \& Roy 1994). In addition, barred galaxies seem to show a 
         flatter gradient than the non-barred ones (Pagel et al. 1979; Alloin et al. 
          1981) for all morphological types. Again, this is not without controversy, 
           because other authors measured large gradients in barred galaxies (Martin \& 
            Roy 1994).

The relation obtained by Zaritsky et al. (1994) between the gradient and the morphological type is quite interesting in the sense that it might indicate that H\,{\sc ii} regions in galaxies of later morphological types, as Sm and Im, should have important differences in metallicity. A similar conclusion was held also by Kunth \& Roy (1996), where they discussed the smoothening of the gradients due to the shear in spiral galaxies, while irregular galaxies, because of their lack of internal movements, might keep their local enrichments. This is contrary to observations, where no significant variations are found in the chemical abundances among the irregular galaxies but only nitrogen local enrichment by Wolf-Rayet stars (see Kobulnicky \& Skillman 1997) and only very few galaxies show differences in the abundances larger than $1\sigma$ (Hidalgo-G\'amez et al. 2001). There are several reasons for the rejection of the possible existence of metallicity gradients in irregular galaxies. The use of the semi-empirical methods in the abundance determination, the large uncertainties in the metallicity values, the small size of the galaxies, etc, have been given for the non acceptance of differences in the metal content throughout an irregular galaxy (e.g. Pilyugin 2001). A recent example is the Local irregular galaxy IC 10, for which Magrini \& Goncalves (2009) obtained differences in the oxygen abundance as large as $0.6$ dex but they still said that ''there is no indications of radial gradients´´. Therefore, there is no irregular galaxy with an accepted gradient or variations in the oxygen abundance. On the contrary, there are several dozens of spiral galaxies where abundance gradients have been determined, some of them as small as $-0.01$ dex/kpc. The main reason argued for the acceptance of such small gradients is the large size of the spiral galaxies, most of the time larger than $10$ kpc. Therefore, although the gradient is small, it gives an important difference between the central part and the outskirts of a certain galaxy. In addition to this situation, there are low-mass, spiral galaxies which does not show any difference in the abundance among the H\'{\sc ii} regions. The best studied one is NGC 1313 (Walsh \& Roy 1997). Moll\'a \& Roy (1999) concluded that the absence of oxygen gradient in NGC 1313 is due to its low mass, which resembles irregular galaxies. To our concern, NGC 1313 is the only late-type spiral galaxy with no gradient, and therefore unique. If a large population of spiral galaxies with no gradient are found they will be the transition between the normal late-type Sm with larger gradient, as suggested by Zaritsky et al. (1994), and the Im galaxies with no gradients. 

The main goal of the present investigation is to check if there are any other late-type, low-mass, spiral galaxy without a metallicity gradient. The abundance gradient of four dwarf spiral (dS) galaxies have been determined in order to check if their gradients increase with the morphological type, as suggested by the relationship by Zaristky et al. (1994) or, on the contrary, they got shallower approaching the distribution abundance in irregular galaxies. All the galaxies in the sample are of low mass, both total and gas, low luminosity and small size, very similar to the characteristics of NGC 1313. Moreover, their H\,{\sc ii} regions are extended and their inclinations are not excessive. none of them have previous spectroscopic studies. For more details about the galaxies in the sample, the reader is referred to Hidalgo-G\'amez et al. (AJ, submitted; hereafter paper I).

In the next section, the determination of the chemical abundance gradients for each of the four galaxies in the sample is presented. In Section 3, a discussion on the accuracy of the values is given. The relationship between these gradients and other global parameters of the galaxies is also discussed. Brief conclusions are presented in Section 4. 

\section{Gradient abundances in dwarf spiral galaxies}

A total of $29$ H\,{\sc ii} regions in four dS galaxies were studied in a previous paper (Hidalgo-G\'amez et al. AJ submitted; hereafter paper I). Two of the galaxies, UGC 5242 and UGC 5296, have a small number of regions while for the other two, UGC 6205 and UGC 6377, the number of regions are larger, of the order of $10$. Although the number of H\,{\sc ii} regions detected in these galaxies is much smaller than in other spiral galaxies, they agree with the observation in other dS galaxies (Hidalgo-G\'amez 2005; Reyes-P\'erez 2009).  The gas mass of all of the galaxies under study is small, about 10$^9$ M$\odot$ (Hidalgo-G\'amez, unpublished), and  similar to the gas content that the low-mass, late-type spiral NGC 1313 (Walsh \& Roy 1996). This galaxy, however is brighter but with a similar chaotic spiral structure than those studied here. 

Two observables are needed in order to determine the abundance gradient: the galactocentric distance of each H\,{\sc ii} and its chemical abundance. In order to obtain a proper distance from the center of the galaxy to the H\,{\sc ii} regions a zero point should be considered. This could be photometric (the place with the maximum surface brightness) or dynamical (the place with the lowest rotation velocity). In the present investigation, the first one is considered. From $V$ and $R$ images (obtained by one of the authors at the 1.5m telescope of San Pedro M\'artir-OAN) a photometric center of the galaxy was determined. To identify the H\,{\sc ii} regions, H$\alpha$ images of each galaxy were compared with the images from the acquisition camera except for UGC 6277 for which no H$\alpha$ image was available. Due to the low number of H\,{\sc ii} regions in this galaxies this was an easy task. Only in UGC 6377, because the lack of the H$\alpha$ image, there might be some problems i
 n the identification. Once the H\,{\sc ii} regions were identified, the H$\alpha$ images were compared with the $V$ and $R$ ones, in order to determine the photometric center in the H$\alpha$ image. A distance was obtained considering the pixel sizes of every CCD involved and the distance to the galaxy, which might be the greatest source of uncertainties. Inclination corrections for each galaxy were applied, using the values tabulated by NED at the $25$ mag~arcsec$^2$. 

The abundances of the H\,{\sc ii} regions studied here were determined in paper I. Here, a brief summary of the 
values are given but the reader is referred to paper I for details. In one H\,{\sc ii} region, the forbidden oxygen line at 4363\AA~ is detected and, therefore the standard method, as described in e.g. Osterbrock 1989, can be used to determine the chemical abundances. In the other H\,{\sc ii} regions, the semi empirical methods are needed. A total of four different semi empirical methods are used. The results are that the abundances are in the range between $8.6$ and $7.7$ dex for most of the H\,{\sc ii} regions. For a study of the feasibility of the semi empirical methods the reader is referred to Hidalgo-G\'amez \& Ram\'{\i}rez-Fuentes (2009). 

In order to study as carefully as possible the gradients in these four galaxies, a gradient value will be determined with each set of abundances obtained in paper I: the R$_{23}$, the $P$, the N$2$, the N$3$ and the average abundances.  In addition, a comparison among the values determined with each method will be done. This is very interesting  because it is believe the bending of the slope observed in some galaxies, as M101 (Zaritsky 1992), is due to an artifact´s in the metallicity determination (Pilyugin 2003).

As discussed in paper I, each method has its own advantages and drawbacks: the R$_{23}$ and $P$ used a larger number of H\,{\sc ii} regions whereas only those H\,{\sc ii} regions with the highest S/N are used the $N2$ and $N3$ methods.  The gradients will be obtained from a least-square fitting to all the abundance values determined for the H\,{\sc ii} regions for each galaxy. A more robust value of the slope could be obtained from a bi-variate fitting: first considering the galactocentric distance as independent and then treating the abundances as the independent variable (Elmergreen et al. 1996). The fitted slope and constant are a combination of the values for each single fitting. However, in this study a single square-fitting has been used because it provides a simple and quick comparison with other investigations. In this statistical approach the inclusion of one single odd point might lead to an important change in the slope and do not reflect the real tendencies (Vila-Co
 stas \& Edmunds 1992). Therefore, and in addition to this mathematical value of the gradient, we have also considered the differences between the innermost and outermost regions (I/O in Table 1) in each galaxy as well as between the most and less metallic ones (L/M in Table 1). Finally, a gradient can be obtained when averaging all the abundances from all the H\,{\sc ii} regions located in a certain range of distance from the center. Therefore, a single abundance is obtained for a single galactocentric distance. The advantage of this last determination is that the possible odd abundances or uncertainties in the galactocentric distance are smoothed out. We advance that the agreement among all the values of the slopes/gradients is remarkable for most of the galaxies.

As said, two galaxies have a very small number of H\,{\sc ii} regions detected and therefore, the gradient values determined might be not very reliable. Indeed, for UGC 5296 and UGC 5242 the total number of H\,{\sc ii} regions is much smaller than the minimum number of data-points needed for a reliable gradient determination according to Dutiful \& Roy (1996). There are only five H\,{\sc ii} regions in UGC 5296 (R\'eyes-P\'erez 2009) and about eight in UGC 5242 (Hidalgo-G\'amez, in preparation). Therefore, we are aware that the gradient values for these galaxies are less confident than for the rest of the sample. Nevertheless, there are some other galaxies where a gradient is determined from only $4$ regions, as NGC 3351, NGC 5068 and NGC 4395 (Vila-Costas \& Edmunds 1992). 

\subsection{UGC 6205}

UGC 6205 is the galaxy with the largest number of H\,{\sc ii} regions detected in our sample: a total of $11$ regions were studied in this investigation. The gradient determined with the semi empirical methods are shown in the first column of Table 1. Also, the values determined using the most and less metallic regions, as well as the most internal and external ones are presented in rows 6 and 7, respectively. Finally, the slope from the averaged distances is presented in row 8 (See below). If the three latter values of the gradients are similar to those obtained with the least-squared fitting might indicate the robustness of the determination. This is the situation for UGC 6205, with all the values of the gradients ranging from $-0.2$ to $-0.4$ dex/kpc. An average value of $-0.31$ dex/kpc with a dispersion of $0.03$ will be considered. 

Another remarkable results presented in Table 1 is the steepness of the gradient, being larger than the values obtained for most of the galaxies studied so far (e.g. see Table 4 in Vila-Costas \& Edmunds 1992). Why? One might think that this galaxy shows very extreme abundance values, but the most external region, located at 3.1 kpc, has a metallicity of $7.7$ dex while the most internal regions have metallicities of $8.7$ dex. The difference is only of $1$ dex, which is not so large. There are other spiral galaxies with such difference in their abundances, or larger, as M81 (Garnett \& Shields 1987). The point here is that M81 is very much larger than UGC 6205. So, UGC 6205 has the same difference in the abundance than other galaxies but the galactocentric distances of the H\,{\sc ii}  are smaller. Therefore, so large gradient. 

The next question to be addressed is which is the region responsible for such a steep gradient. One might think that those regions with low S/N are responsible for the slope because their abundance value might not be so reliable. This is not true for several reasons: firstly, the gradients obtained with the $N2$ and $N3$ methods, which included only those high S/N regions, give only slightly shallower gradients. Moreover, the values determined without the least-squared fitting are of the same order except when the gradient is determined from the most and less metallic regions which is of $-1.1$ dex/kpc, and it will be ignored  because of its doubtful meaning.   

Another way to check the reliability of the steep gradient, is, as previously said, using an average abundance for different ranges of galactocentric distances: $0$-$0.6$ kpc, $0.6$-$1.2$ kpc, $1.2$-$1.8$ kpc, $1.8$-$2.4$ kpc, $2.4$-$3.0$ kpc, and $3.0$-$3.600$ kpc. As said, the variations of the abundance are smoothed and the low S/N regions will not be the responsible of the value of the gradient. These are shown in the eighth row of Table 1 and it is about $-0.3$ dex/kpc for this galaxy. Whatever the method is used for the gradient determination, will end up in a very high value. Then, it can be concluded that such value might be real. 

\begin{figure}
\centering
\includegraphics[width=14cm]{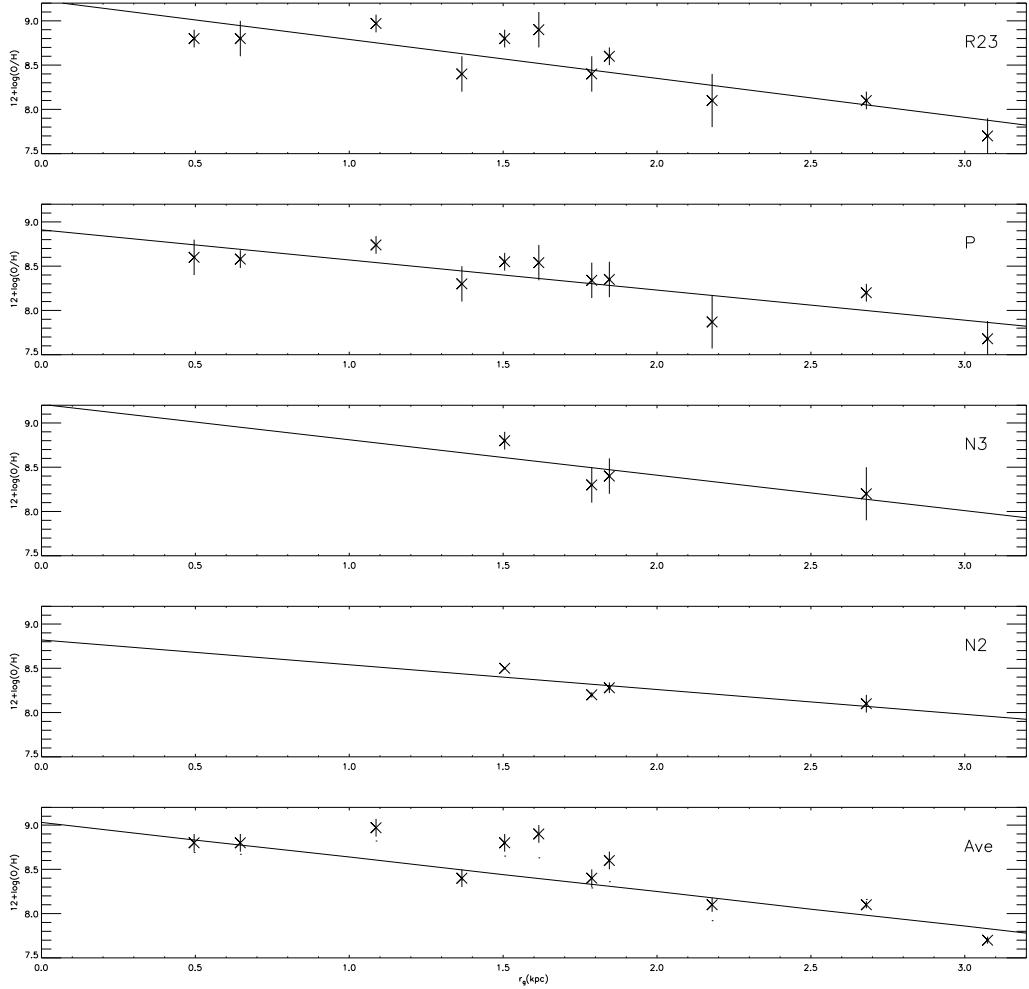}
\caption{The abundances along the galactocentric radius for UGC 6205, determined with the four methods described in paper I. From top to bottom are the R$_{23}$, the $P$ method, the $N3$, the $N2$ and the mean metallicity from these four methods (See paper I for details on each method and the abundance determination). The solid line is the least-squared fitting to all the H\,{\sc ii} regions. Errobars in the abundances are also shown.} 
\label{fig1}
\end{figure}

Figure ~\ref{fig1} shows the metallicity vs. the galactocentric radius, in kpc, for each of the five metallicity determinations. The errorbars in the metallicity and in the galactocentric distances are shown but the latter are smaller than the symbol. The solid line is the fitting using all the H\,{\sc ii} regions. The fitting is, in general, very good with all the methods except for regions U05a1, U05c1, and U05b5 which have odd abundance values for their galactocentric distances. If those regions are not considered, the gradients do not change significantly. 

It can be noticed in figure ~\ref{fig1} that the gradient seems to change along the radius of the galaxy, with a plateau for radii up to $1.6$ kpc, whatever method is used for the abundance determination. This is motivated mainly because the high metallicity of regions U05c1 and U05a1, which as stood in paper I, they might be something more than simply H\,{\sc ii} regions. In any case, an internal (r $<$ 1.4 kpc) and external gradients were determined without these three anomalous regions, and no differences were found. 

It can be concluded that the gradient in this small spiral galaxy is very steep, of about $-0.3$ dex/kpc, with no dependence on the metallicity method and no differences along the galactocentric distance.

\subsection{UGC 6377}

As mentioned in paper I, two of the nine H\,{\sc ii} regions in this galaxy have abnormal abundances and they might be planetary nebulae instead of normal H\,{\sc ii} regions. Therefore, they are not going to be considered in the determination of the gradients. We will use the other seven regions in order to obtain the gradient. The values from the $N2$ and $N3$ abundances have only two regions and they are located very close each other, thus no reliable values can be obtained. The values of the slopes are shown in the second column of Table 1.  It can be seen that the $R_{23}$, the $P$ method and the average abundance gradients are very similar, from $-0.15$ to $-0.20$ dex/kpc. Also, those values are very similar to the slopes determined from the most external/internal regions, with values ranging from $-0.16$ to $-0.13$ dex/kpc. The slopes determined from the most/less metallic regions are larger, of the order of $-0.3$ dex/kpc. 
The main reason is that the less metallic region, U77c3, is not the most external one. The abundance differences in UGC 6377 are lower than for UGC 6205, only $0.8$ dex, but the distances considered are also smaller, only $2.4$ kpc. Therefore, the gradients are also quite large. 

A reason for such steep slopes are needed. Similarly as in UGC 6205, one might think that the low S/N regions are changing the slope. Only three of the seven H\,{\sc ii} regions used are of high S/N. When only those three are used (with all the caution due to the low number of data-points) the gradient is $-0.11$ dex/kpc, which is more similar to other late-type spirals galaxies (see next section). Another way to check the influence of the low S/N regions is using an averaged metallicity value for certain distance ranges, as for UGC 6205. In this case, those regions with low S/N are not so critical. The distance ranges were $0$-$0.6$ kpc, $0.6$-$1.2$ kpc, $1.2$-$1.8$ kpc, and $1.8$-$2.4$ kpc. The slope in this case is of the order of $-0.23$ dex/kpc. The (weighted) mean gradient is of $-0.2$ dex/kpc, with a dispersion of $0.02$.

\begin{figure}
\centering
\includegraphics[width=14cm]{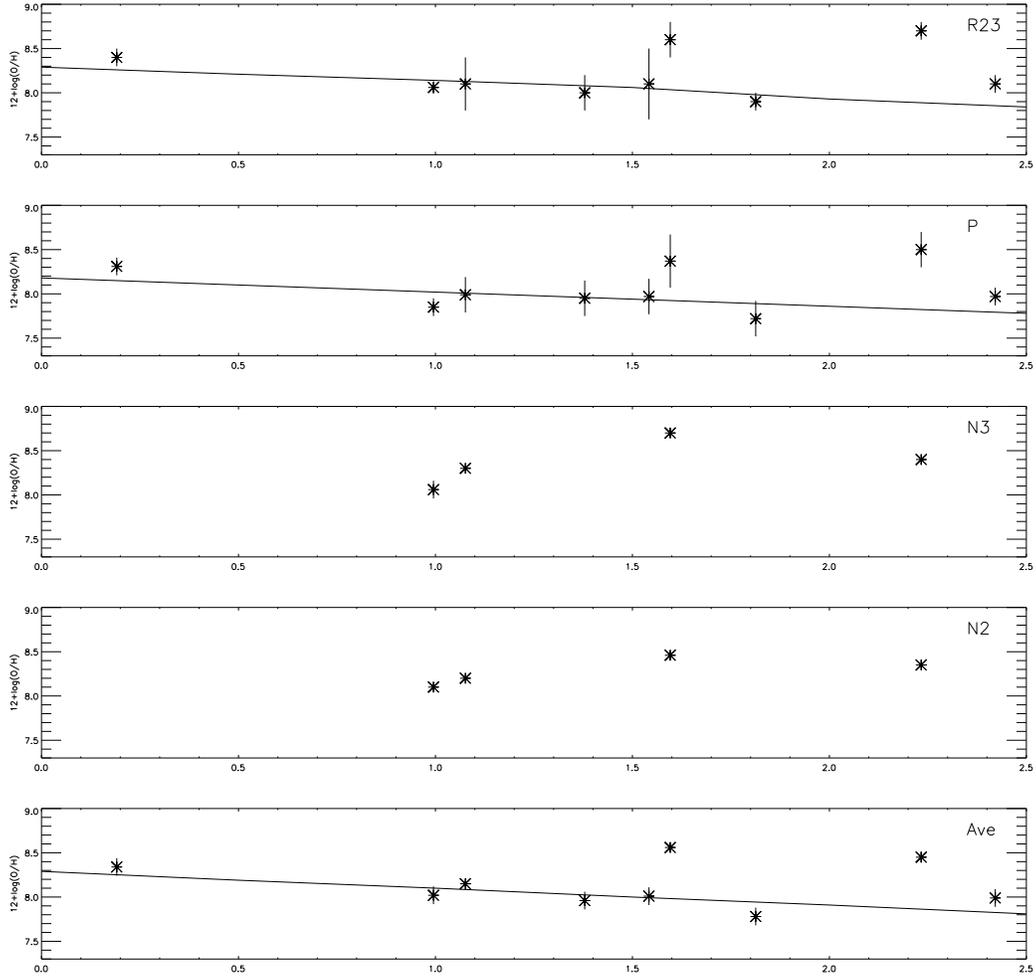}
\caption{The abundances along the galactocentric radius for UGC 6377. Symbols and lines as in figure ~\ref{fig1}. The abundances determined with the $N2$ and the $N3$ methods are not shown due to a low number of data-points (see text). The high-metalliticy abundances of U77a1 and U77b1 are plotted in order to see if they fits the gradient, although they were not considered in the abundance gradient determination. } 
\label{fig2}
 \end{figure}

Figure ~\ref{fig2} shows the plot of the metallicity vs. the galactocentric distance along with the fitting for all the metallicity methods for which a gradient has been determined. The fitting is very good for all the cases. 

From a close inspection of Figure ~\ref{fig2}, one might say that there is a change in the slope of the gradient, being steeper for the internal regions than for the external ones. This situation has been observed in other galaxies such as M101 (Zaritsky 1992) or NGC 2403 (Garnett et al. 1999). Pilyugin (2003) has argued that such bends in the slope of the abundances are artifacts in the abundance determinations. This explanation is not valid here because the flattening of the slope is observed when the abundances are determined with the $P$ method. Considering only those regions most internal than $1.2$ kpc, the gradient is quite steep, of about $-0.35$ dex/kpc,  while the value determined with the outer regions is almost flat. This situation is forced by the abundance of U77c1, which is the most external region but its abundance is high. Without this region, the outer gradient is of the order of $-0.15$ dex/kpc and the difference with the internal regions is not that high. 

Finally, we will discuss about the abundances of U05a1 and U05b1. They have very high abundances for their galactocentric distances. If the low abundance values are considered instead of the high branch ones, they both fit very well the gradient determined by the rest of the data-points. Actually, the value of the gradient does not change at all. So, as concluded in paper I, they are likely PNs embedded in low metallicity H\,{\sc ii} regions.

\subsection{UGC 5296}

There are only four H\,{\sc ii} regions for gradient determination in this tiny galaxy. One might think that no gradient can be determined with such a few number of data-points. But there are at least another three galaxies with values determined with only four metallicity determination. Nevertheless, the confidence of the gradient obtained is not as good as for the two galaxies previously studied.  

The H\,{\sc ii} regions observed in UGC 5296 are located in a line from south to north. For each of these regions, two set of intensities are obtained, as explained in paper I. For those lines obtained from ALICE-MIDAS, the gradient with the $N2$ and $N3$ methods cannot be determined because only two regions are considered. Moreover, the slopes determined with $N2$ and $N3$ VISTA-abundances are positive. This is an indication that the VISTA nitrogen intensities are not very trustworthy. Therefore, no further consideration will be given to those abundances and gradients and only the $P$- and $R_{23}$-abundances determinations will be discussed.

\begin{figure}
\centering
\includegraphics[width=14cm]{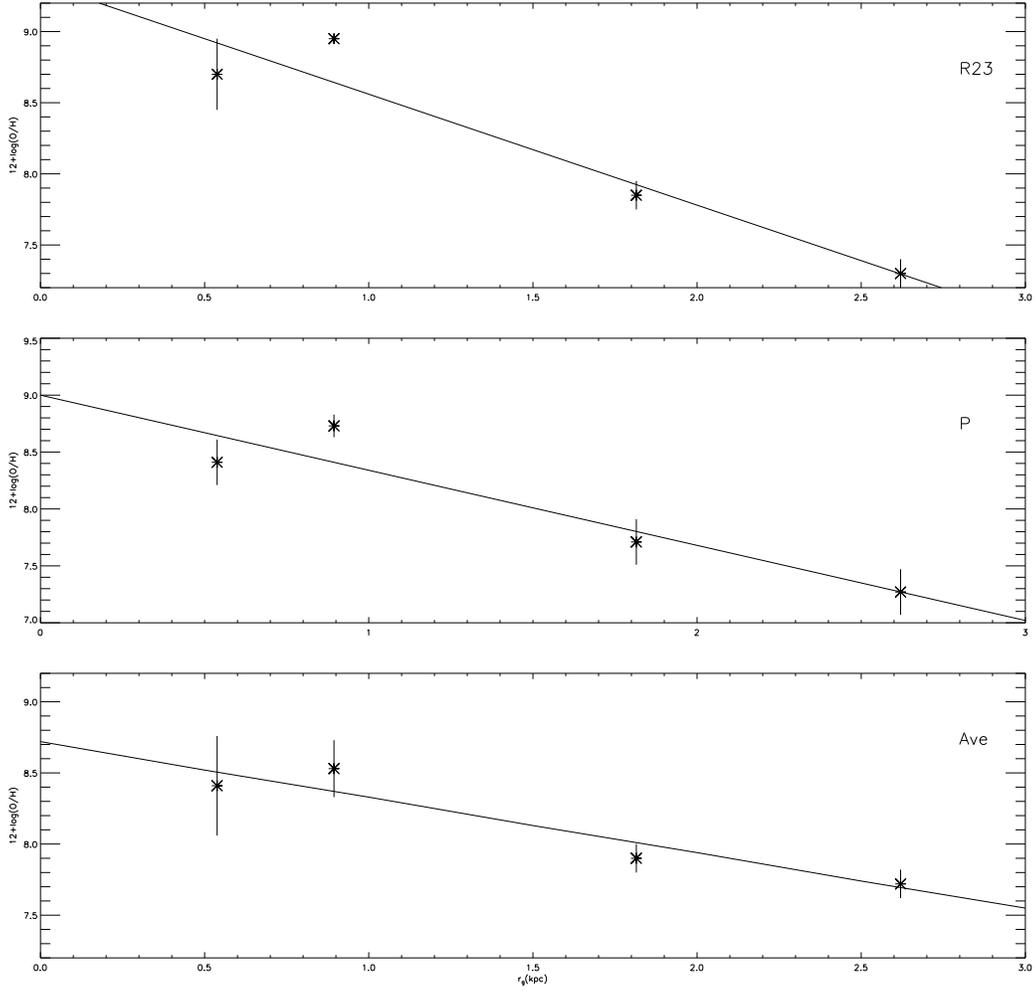}
\caption{The abundances along the galactocentric radius for UGC 5296. Only those abundances determined with the $R_{23}$ (top panel), the $P$ (central panel) methods or the average abundance (bottom panel) are shown, while the abundances determined with the $N2$ and the $N3$ methods are not shown due to a low number of data-points (see text for details). Symbols and lines as in figure ~\ref{fig1}.} 
\label{fig3}
 \end{figure}

In order to obtain a more reliable determination of the gradient, a single value of the abundances is obtained for each H\,{\sc ii} region by averaging the values from the VISTA data-set and the ALICE-MIDAS one. This can be done because the abundances between these two data-set are very similar. The gradients determined in this way are shown in column 3 of Table 1 and in Figure ~\ref{fig3}.  The value from the average abundance is $-0.6$ dex/kpc, which is quite unrealistic as well as those determined from the $R{23}$ and $P$ methods which are even steeper. The reason for such a steep value is the large difference in metallicity among the internal and the external regions in this galaxy, $1.2$ dex, and the small galactocentric distance range, $2$ kpc. Although a gradient at fixed distance has little sense because it will be determined with only two regions, it is useful because it shows that the abundances change dramatically, from $8.5$ dex at the center to $7.7$ dex at the o
 utskirts.

The slope derived using the more/less metallic regions is $-0.6$ dex/kpc while the value using the out most/inmost regions is $-0.3$ dex/kpc. These values are similar to those determined with the mathematical fitting. This is because U96a2 is probably over enriched as discussed in paper I. Such enrichment does not force the steep gradient, because a value of $-0.52$ dex/kpc is obtained without this region. One might think that region U56a1, with its very low metallicity, is forcing such steep gradient. Actually, the $R_{23}$ and $P$ gradients do not change when this region is not considered. On the contrary, the average metallicity gradient does shallow mainly due to the higher abundances for U96a4. 

Values as large as $-0.6$ dex/kpc are not acceptable because they are quite unrealistic. The value of the gradient adopted for this galaxy is $-0.4$ dex/kpc, as stated in Table 1. The dispersion is about $0.1$. Such gradient is large but similar to the one of UGC 6205. From  the results of the averaged distances determination discussed above, it can be concluded there is a real, and large, change in the abundances of this galaxy. Such differences cannot be explained by differences in the SFR (R\'eyes-P\'erez 2009). Finally, it is important to point out that a more accurate determination of the gradient will be very difficult, because there is only one more H\,{\sc ii} region in the galaxy. In any case, more high quality data  are needed in order to confirm these results.

\subsection{UGC 5242}

The number of H\,{\sc ii} regions is also small in this barred galaxy, but for the same reasons as for UGC 5296, a gradient will be determined. Moreover, as being the only barred galaxy in our sample, it will be very interesting to know if the gradient is similar to the other barred late-type galaxies. The oxygen abundances can be determined for only five regions. Also, as for UGC 5296, there are two set of data for each region. Therefore, a total of eight different estimations of the gradient can be obtained for this galaxy. Those values determined with the ALICE-MIDAS abundances are very similar, but those obtained from the VISTA abundances are not. Two of them are positive. The best way to proceed is, just as in UGC 5296, to determine an average abundance for each H\,{\sc ii} region from those obtained with ALICE-MIDAS and with VISTA data-set of intensities. In this case, the values of the gradient vary from $-0.006$ dex/kpc to $-0.18$ dex/kpc (see Table 1). These values a
 re slightly larger than those determined for other barred late-type galaxies. 

Again, the values of the slope obtained with the most/less external regions are very similar to those obtained with the fitting. On the contrary, the slope determined from the most/less metallic regions are very large. This value could not be realistic and the main reason for such a steep gradient is that the less metallic region, U42c2, is not the most external one. The differences in abundances between the regions is of almost $0.5$ dex but with less than $1$ kpc in separation. Moreover, the most metallic region has the lowest S/N. When another region is considered, a gradient of $-0.15$ dex/kpc is obtained, which is more similar to the rest of the values. 

Although it should be taken with care due to the small number of region involved, a metallicity at averaged distances can be obtained as well as a gradient. Values from $-0.02$ to $-0.14$ dex/kpc are obtained, which are only slightly smaller than those obtained with all the regions. An average gradient of $-0.17$ is obtained from all the values in Table 1, with a dispersion of $0.06$. 

\section{Discussion}

\subsection{How accurate these gradients are?}

In a investigation like this there could be a lot of sources of uncertainties. 
Some of them could be related to uncertainties in the distance determination to the galaxy itself. Normally, distances are very difficult to obtain when no primary candles are considered. This is the case for all the galaxies in this sample. The distance used here were those from Hidalgo-G\'amez (2004), normally based on secondary indicators.  Distance determinations can change the value of the gradient if the new determination differs greatly from the old one. In general, the distances provided by NED/NASA are very similar to those reported in Hidalgo-G\'amez (2004) for the galaxies in the sample. The only important difference is on the distance of UGC 5296, and the gradient changes from $-0.6$ dex/kpc to $-0.45$ dex/kpc, meanwhile for the rest of the galaxies the differences in the gradient determination are smaller than the uncertainties. Therefore, the steep gradients in these galaxies are not due to wrong distance determinations. Also, the galactocentric distances might 
 have influence on the slope of the abundance, but not in this case as indicated by the results of the gradient determined at fixed distances.

Probably, the largest source of uncertainty in the slopes obtained in this investigation is the abundance determination. As they cannot be calculated with the standard method (e.g. Osterbrock 1989), the abundance value itself might be very uncertain. As described in paper I, four different semi empirical methods were used in the abundance determination. Moreover, a weighted-average value of the abundance was obtained for each region. Therefore, the gradients determined with such metallicity should be very reliable. Moreover, those regions which are suspected not to be a normal H\,{\sc ii} region are not considered in the gradient determination, as in UGC 6377. There are some regions in UGC 6205, as U05b4 and U05c3, which were considered of high metallicity despite the small value of the log(N/O) ratio (section 3.2 in paper I). Nevertheless, there is not change in the slope because they are located at the intermediate part of the galaxy, and therefore their inclusion is not cru
 cial. As discussed in paper I, despite the uncertainties in the metallicity determination, it is not quite likely that a high metallicity region was misclassified as of low metallicity one. 

Another source of uncertainty might be the low number of measurements for the determination of the gradient. Dutil \& Roy (2001) said that in order to obtain a robust measurement of the gradient at least $16$ data-points are needed. This is an impossible achievement for the dwarf spiral galaxies, mainly because most of them do not have such a large number of H\,{\sc ii} regions. From the study of Reyes-P\'erez (2009) and the H$\alpha$ images obtained for two dozens of dS galaxies by Hidalgo-G\'amez, it can be concluded that less than $10$ dwarf spirals out of more than $100$ have more than $15$ H\,{\sc ii} regions. In particular, all the regions were studied for UGC 6205 while only one and two more exist for UGC 5296 and UGC 5242, respectively (Reyes-P\'erez \& Hidalgo-G\'amez, in preparation). Therefore, the results here should be taken with care because of the small number of data-points, but it has to be understood that there are no more regions to be used. Finally, it has
  to be said that the gradients determined for another 11 galaxies in the literature were determined with only five measurements or less (see table 4 in Vila-Costas \& Edmunds 1992). 

Dutil \& Roy (2001) also said that not only the slope is uncertain due to the poor sampling, but also the gradient might change when more data-points are added. This can be studied here comparing the gradients determined with the $R_{23}$ or the $P$ methods and those determined with the $N_2$ or $N_3$ ones, because the number of H\,{\sc ii} regions involved are different, being the latter much lower. The best galaxy is UGC 6205 
because the number of abundance determinations are not low even with the $N_2$ and $N_3$ methods. In Table 1 it could be seen that there are differences in the gradient if determined with $11$ measurements or with four, but they are small. One might think that this is the situation for both UGC 5296 and UGC 5242 in the sense that there are important differences in the gradients determined with the different methods. For these galaxies the number of data-points are the same with both methods and it is a problem of false nitrogen detections. Therefore, it can be concluded that none of the possible source of uncertainties have a real influence on the gradient of the abundances. 

A final criticism that can be made to the data presented here is that as the abundances are determined with few lines, they might be not very reliable. This should be always taken into account when working with these galaxies, as well as with other galaxies for which the abundances have been determined from the [OIII]/[NII] or the [NII]/H$\alpha$ ratios, in the local Universe (e.g. Roy et al. 1996) or for high $z$ galaxies (e.g. Schulte-Ladbeck et al. 2004). As can be seen from Table 1 the value of the gradients are, in general, very similar for each galaxy when the abundances are determined with different methods. Moreover, the values determined with the less/most metallic regions or the internal/external regions are also very similar to those determined from the least-squared fitting.  Therefore, we think that, although the abundances for some particular regions can be somehow uncertain, the gradients obtained here are very robust. 

\subsection{How real these gradients are?}

There are few tests that can be done in order to see if the values obtained here are real or due to some artifacts related to the uncertainties in the abundance determinations. The first one is to determine the gradient, if any, in the [OIII]/H$\beta$ ratio. Differences in this ratio were detected in the pioneering work of Searle (1971). Actually, he considered that those  were due to differences in the abundances instead of in the excitation. This is shown in Figure ~\ref{fig5} for those non-barred galaxies in the sample. Along with the data-points and their errorbars, the least-squared fitting is shown (solid line). Only  UGC 6205 shows a positive, and steep, gradient (larger ionisation for the most external regions). The dotted lines show the locus where the data-points can be located if the line intensities change by their uncertainties. All but two regions lay inside these lines. On the contrary,  UGC 6377 shows a very unexpected behaviour, with a negative and steep gradien
 t. This is due to the high ionisation of U77b2, and the low ionisation of U77a1, both with low S/N. Considering only those H\,{\sc ii} regions with medium and high S/N, the gradient is positive and very steep, of $-0.53$ dex/kpc, which is shown by the dashed line. Finally, the excitation gradient is almost flat for UGC 5296 (0.11 dex/kpc), with a high value for the most internal region and a low excitation for U96a1.
The other gradient that can be determined is the [NII]/H$\alpha$ ratio. The only galaxy for which is it meaningful is UGC 6205. The nitrogen gradient is also very steep in UGC 6205, of $-0.44$ dex/kpc. From both line ratios, it can be concluded that the oxygen gradients determined are real, especially for UGC 6205 for which both ratios give very steep gradients. 

Finally, a total gradient for non-barred dwarf spiral galaxies can be determined. Figure ~\ref{fig6}a shows the abundances, from the average values, vs. the optical radius, in kiloparsecs, for the three galaxies studied here while the average abundances at a fixed distance for all the H\,{\sc ii} regions of the three non-barred galaxies is shown in figure ~\ref{fig6}b. Those regions which are not considered in the gradient determination for a particular galaxy are also not considered in the fitting here but they are shown for completeness. Although the dispersion is large, particularly at intermediate distance, the gradient is also very steep, of $-0.3$ dex/kpc.  Figure ~\ref{fig6}c shows the same abundances but to the normalised distance (r/r$_{25}$). In this case, the slope of the fitting is the same, $-0.3$. Two important conclusions can be obtained from this figure. Firstly, the H\,{\sc ii} regions of three dS galaxies are non-distinguishable in the plots, in the sense that
  all of them have the largest metallicity for the central regions and the lowest for the most external ones. This is more clear in figure ~\ref{fig6}c, with a very low dispersion. The second conclusion is that these three dS galaxies have an steep abundance gradient, being the gradient obtained from the abundances at a averaged distance, show in figure ~\ref{fig6}b, the most robust. Therefore, it seems that the oxygen gradient for dwarf late-type spiral galaxies might be larger than for their normal-size counterpart. A larger sample of dwarf spiral galaxies are needed for a more conclusive result.   

\subsection{Is there any dependence of the gradient with other parameters?}

Vila-Costas \& Edmunds (1992) found out that the gradient is related with the total mass of the galaxy for non-barred galaxies and any morphological type. They also obtained correlations with the absolute magnitude, when all the morphological types and barred galaxies are included. Also, Zaritsky et al. (1994) found out a good correlation between the gradient and the morphological type, with larger slopes for late spirals. 

Here, we are going to explore the relationship among the gradient and other characteristics for late-type spirals only. In order to do this, information on the optical radius, surface brightness, absolute magnitude and gas mass and density has been obtained for five late-type (Sd or later), non-barred spiral galaxies and four barred galaxies from different authors, as Webster et al. (1983), McCall et al. (1985), Zaritsky et al. (1994), van Zee et al. (1997) and Hidalgo-G\'amez (2004) and references therein. None of these parameters except the absolute magnitude were studied by Vila-Costas \& Edmunds (1992) or Zaritsky et al. (1994). The results are shown in Figures from ~\ref{fig7}a to ~\ref{fig10}a for non-barred galaxies and in Figures from ~\ref{fig7}b to ~\ref{fig10}b for barred ones. In these figures, the stars correspond to the galaxies in the literature and triangles represent the galaxies in this study (dwarf spiral galaxies). Also, the solid line is the fitting to th
 e non-dwarf galaxies sample while the dotted line is the fitting to all the galaxies, including the dwarf ones. The regression coefficients of the former are shown at the top-right corner of each figure. 

The largest value of the regression coefficient for non-barred galaxies are those for the M$_b$ and the optical radius, indicating that the gradient of non-dwarf, late-type spirals shows a trend with these two parameters.  There is also a weak correlation with the gas mass (total r$_g$ = $-0.61$) but not trend at all for the gas surface density, with a regression coefficient of $-0.3$. When dwarf spirals  are included, the regression coefficients do not change much, although the slopes became shallower for all the cases.  There might be several reasons for the small regression coefficient in Figure ~\ref{fig10}a as a much smaller number of data-points. Also, as the values of the dS are determined averaging the gas mass over the area of a disc and not considering an specific disc model, there must be differences between the dS and the Sm gas surface density. In any case, the value obtained for UGC 6205 seems to be similar to the other five Sm galaxies, but not UGC 5296 where 
 the surface density is much smaller. In order to fit the non-dwarf trends, both UGC 6205 and UGC 5296 might have lower gradients, between $-0.15$ and $-0.30$ dex/kcp. They are, in any case, larger than those values for non-dwarf galaxies. They both also seem to have very low content of the gas mass.  

Figures ~\ref{fig7}b to ~\ref{fig10}b show the same as Figures ~\ref{fig7}a to ~\ref{fig10}a, but for barred galaxies, with a total of four galaxies from the literature (NGC 4395, NGC 925, NGC 1313, and NGC 5068) and UGC 5242 (triangle). Now, the absolute magnitude 
(r$_g$ = $-0.93$) and the gas surface density (r$_g$ = $-0.73$) show a correlation with the gradient. 

The results presented here seem to be in agreement with previous investigations. The correlation of the gradients with both, the radius and the absolute magnitude might be related with the correlation with the morphological type. The late-type galaxies tend to be smaller and with larger magnitudes than earlier galaxies (Hidalgo-G\'amez 2004; van der Bergh 2008). Moreover, barred galaxies tend to be larger than non-barred galaxies of the same M$_b$ (Hidalgo-G\'amez 2004). The main difference resides in the gas density surface, with a strong correlation for barred galaxies, while none for non-barred. The reason for such difference is not well understood yet. It might be related with some dynamical processes of the bar acting on the
gas mass distribution. Or it is just a problem of poor sampling with a small range in values of the gradient.

Figures 7-10 can also be used to check the reliability of the gradient of the individual galaxies. Adopting the definition of dwarf spiral galaxy by Hidalgo-G\'amez (2004) and Figures 7a and 8a, dwarf spirals might show slopes larger than $-0.15$ dex/kpc, which are in agreement with the results presented here. As said above, the values for UGC 6205 and UGC 5296 seem to be too large. Gradients of about $-0.2$ dex/kpc for UGC 6205 and between $-0.2$ and $-0.3$ dex/kpc for UGC 5296 might fit all the correlations presented here while, considering the uncertainties in the gradient, the value for UGC 6377 seems to be quite good. The fitting of the only barred galaxy in our sample, UGC 5242, is very good for the relationship with the M$_b$ and the M(H\,{\sc i}), while the gas surface density is very large for its gradient.  Therefore, it can be concluded that despite all the uncertainties and caveats, the gradient for these four particular dwarf spirals are really quite steep and th
 ey follow the same relationship as late-type spirals; that is,  smaller, and less brighter galaxies have larger gradients, regardless of their gas mass.

\section{Conclusions}

The abundance gradient for four dwarf spiral galaxies have been obtained, none of them previously studied. Contrary to expected, the three non-barred galaxies show very steep gradients, larger than $-0.2$ dex/kpc, while the barred galaxy UGC 4252 shows a shallower one, of only $-0.10$ dex/kpc. Therefore, it seems that barred, dwarf galaxies have smaller slopes than non-barred ones, as previously discussed. Although the gradients look very steep compared with the values for the Milky Way, they follow the trend defined by other late-type, non-dwarf galaxies. The gradients for UGC 6205 and UGC 5296 are, at least, of $-0.2$ and $-0.3$ dex/kpc, respectively. In order to obtain a conclusive answer, more late-type, dwarf and non-dwarf are needed.

The increase of the slope with Hubble type has been reported before and it is confirmed  with these new gradients of very small spiral galaxies. There are indications than the dwarfer the galaxy is, the steeper the gradient. This might be a very important conclusion. As concluded in Hidalgo-G\'amez (2004) it seems that dwarf spiral galaxies do not share the same properties as their normal-size counterparts. Actually, their luminosity functions and star formation processes resemble more those for irregular galaxies (Reyes-P\'erez 2009; Reyes-P\'erez \& Hidalgo-G\'amez, in preparation). Therefore, it would be very interesting to ask  why normal irregular galaxies, which are similar to dwarf Sm galaxies in many other aspects, do not have strong gradients in oxygen abundance?

\begin{acknowledgements}

The authors thank J.M. V\,{\'i}lchez, E. Terlevich and A. D\'{\i}az for interesting discussions on the questions discussed on the paper. This investigation is part of the Master Thesis of Daniel Ram\'{\i}rez Fuentes. 
This investigation was supported by SIP20100225 and CONACyT CB2006-60526. This research has made use of the NASA/IPAC Extragalactic Database (NED) which is operated by the Jet Propulsion Laboratory, California Institute of Technology, under contract with the National Aeronautics and Space Administration.

\end{acknowledgements}

{}

\begin{table}
\caption[]{Gradient abundance of the galaxies here. In columns 2, 3, 4 and 5 the name of the galaxy is given, while column 1 presents the different methods used in the gradient determination. The first row gives the gradients when the abundance is obtained with the $R_{23}$ method. The gradients determined with the $P$ abundances are presented in the second row. Rows 3 and 4 give the gradients obtained with the $N2$ and $N3$ methods, respectively. Row 5 shows the gradient obtained with the average value of the metallicity. Also,  the difference in abundance and in distance between the most and the less metallic (L/M) regions and between the most internal and external regions (I/O) in the sample are presented in rows 6 and 7. Finally, the gradient obtained when all the abundance of all regions at a distance bin are averaged is presented in row 8 while the preferred value of the gradient is presented in row 9 with a dispersion value from all of the gradient determinations.  All the values are in dex/kpc. See text for more details.  }
\vspace{0.05cm}
\begin{center}
\begin{tabular}{c c c c c}
\hline
{Method} & {UGC 6205} & {UGC 6377} & {UGC 5296} & {UGC 5242}      \\
\hline
{Z$_{23}$}          & {-0.4} & {-0.15} & {-0.8}  & {-0.06}  \\
{Z$_P$}             & {-0.3} & {-0.16} & {-0.7}  & {-0.006}\\
{Z$_{N2}$}          & {-0.2} & { - }   & {-}     & {-0.2}\\
{Z$_{N3}$}          & {-0.2} & { - }   & {-}     & {-0.14}\\
{Z$_{ave}$}         & {-0.3} & {-0.2}  & {-0.6}  & {-0.1}\\
{L/M metallic}      & {-0.4} & {-0.3}  & {-0.6}  & {-0.6}\\
{I/O regions}       & {-0.4} & {-0.16} & {-0.33} & {-0.15}\\
{averaged distance} & {-0.3} & {-0.25} & {-0.5}  & {-0.07}\\
{gradient}          & {-0.3$\pm$0.03} & {-0.2$\pm$0.02}  & {-0.43$\pm$0.1} & {-0.17$\pm$0.06}\\

\hline
\end{tabular}
\end{center}
\end{table}

\clearpage

\clearpage

\clearpage

\clearpage
\begin{figure}
\centering
\includegraphics[width=14cm]{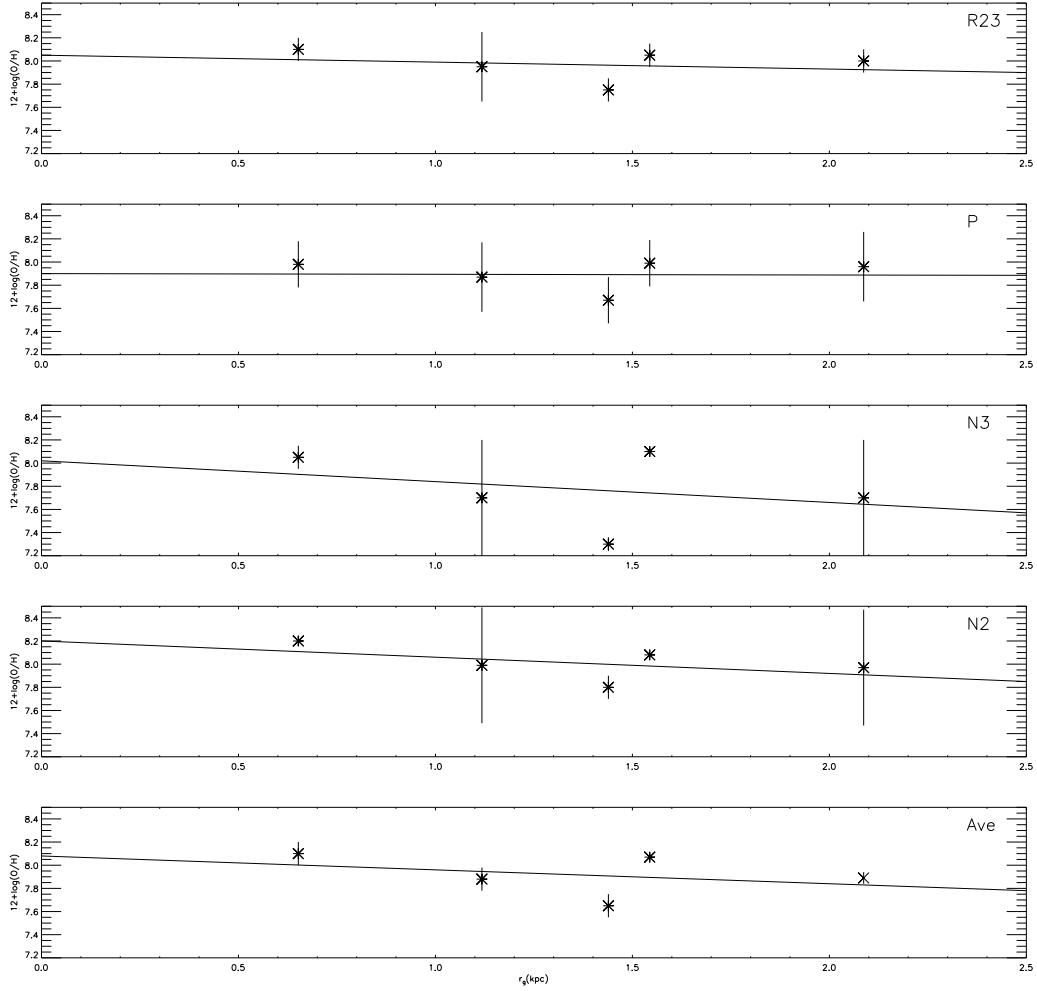}
\caption{The abundances along the galactocentric radius for UGC 5242. Symbols and lines as in figure ~\ref{fig1}. Again, only the abundances determined with the $R_{23}$ (top panel), the $P$ (central panel) methods and the average abundance (bottom panel) are shown.} 
\label{fig4}
 \end{figure}

\clearpage
\begin{figure}
\centering
\includegraphics[width=14cm]{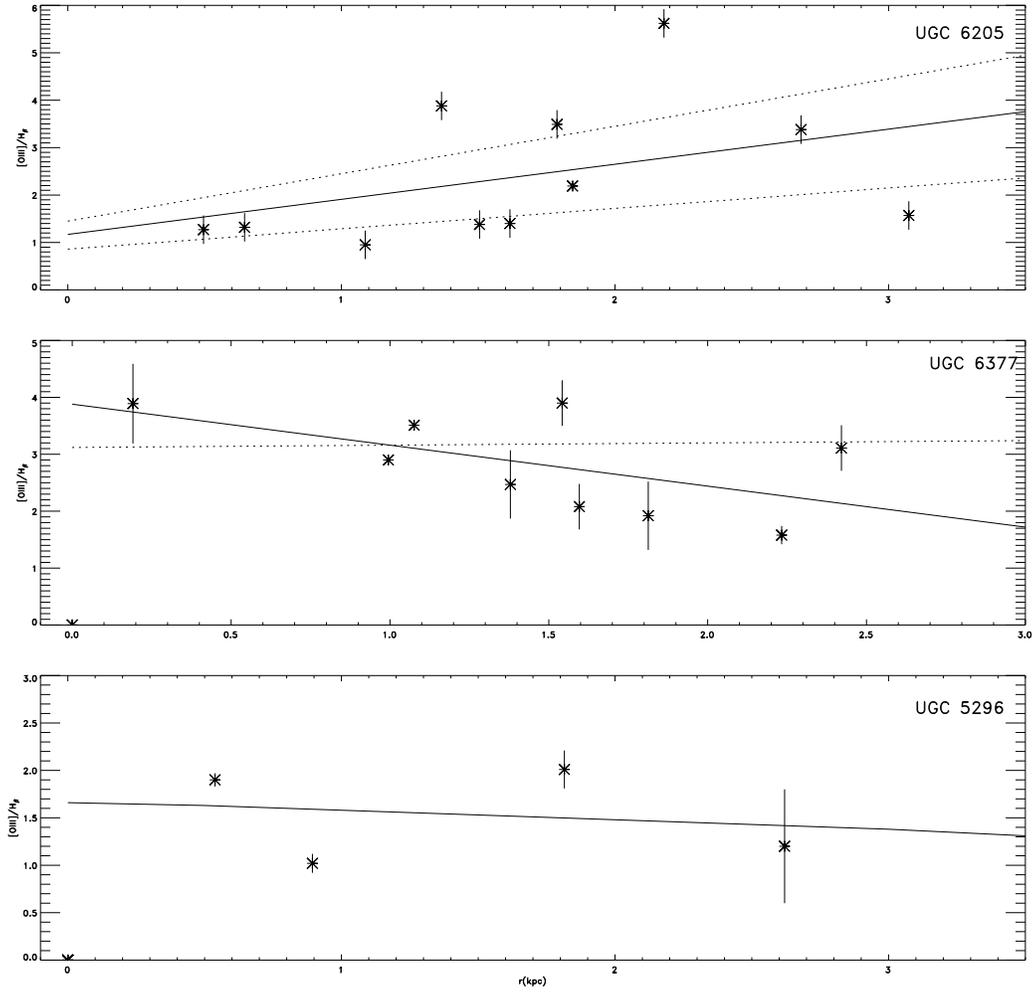}
\caption{The excitation gradients for the non-barred galaxies in the sample. The solid line is the least-squared fitting, while the dotted one in the top panel comprises the locus where the data-points can be located. The dashed line in the central panel is the fitting to the high S/N regions in UGC 6377 (see text for details).  } 
\label{fig5}
 \end{figure}

\clearpage
\begin{figure}
\centering
\includegraphics[width=14cm]{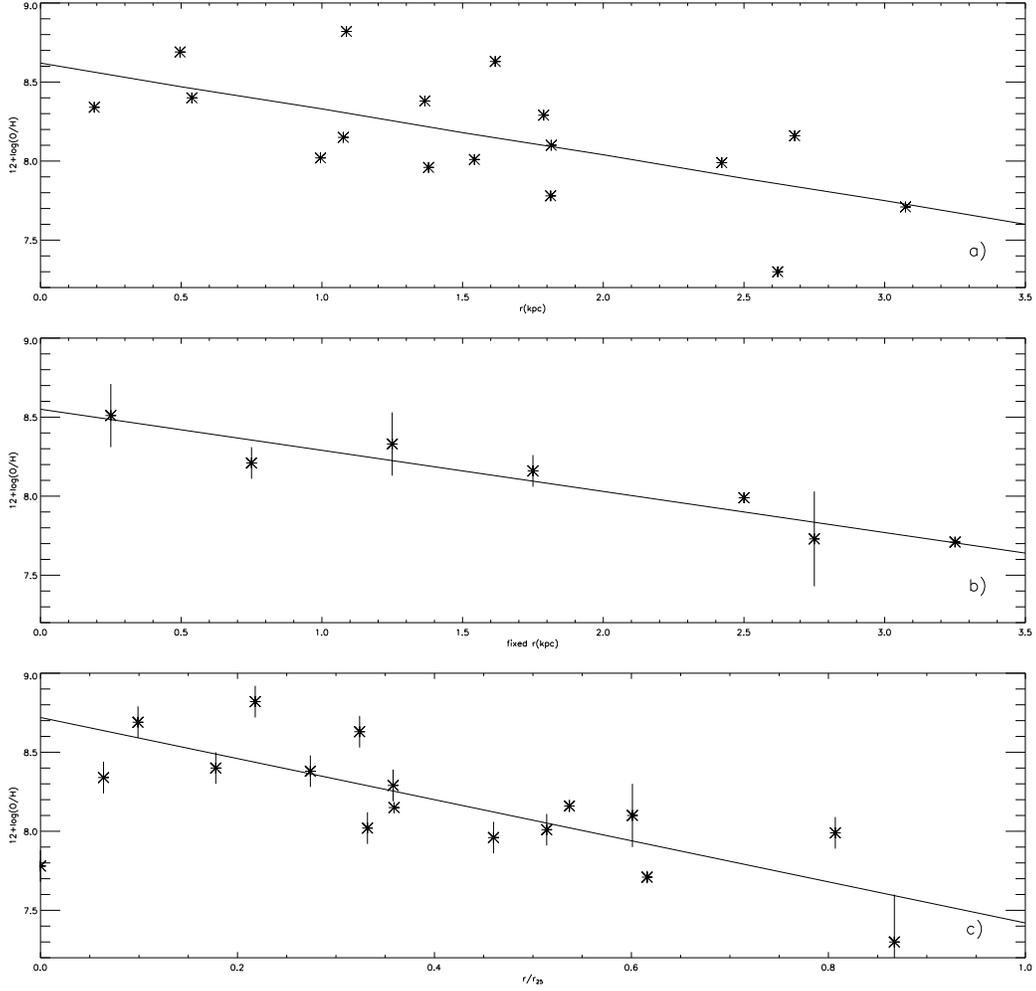}
\caption{The abundance gradient for all H\,{\sc ii} regions for the non-barred galaxies in the sample. In the top panel (a)) the distance is in kpc while the gradient to a normalised distance is shown in the bottom panel. Figure 6b) shows the abundances vs. a fixed distance for all the data-points in the non-barred galaxies.} 
\label{fig6}
\end{figure}

\clearpage
\begin{figure}
\centering
\includegraphics[width=14cm]{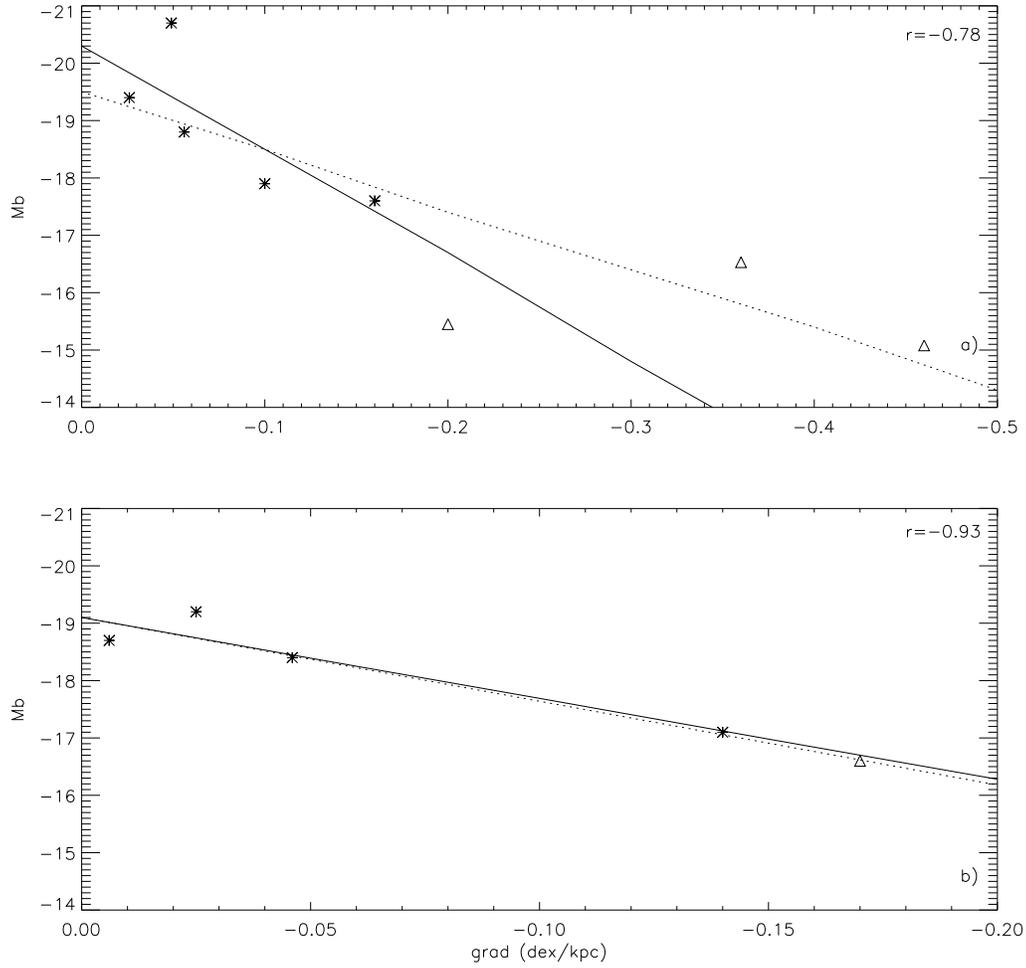}
\caption{The relationship between the absolute magnitude and the gradients for non-barred (a) and barred (b), late-type galaxies. Stars correspond to normal-size Sm spirals while triangles stand for dwarf spirals. The solid line is the fitting to the Sm spirals only, while the dotted line is the fitting to all the data-points in the plot. } 
\label{fig7}
 \end{figure}

\clearpage
\begin{figure}
\centering
\includegraphics[width=14cm]{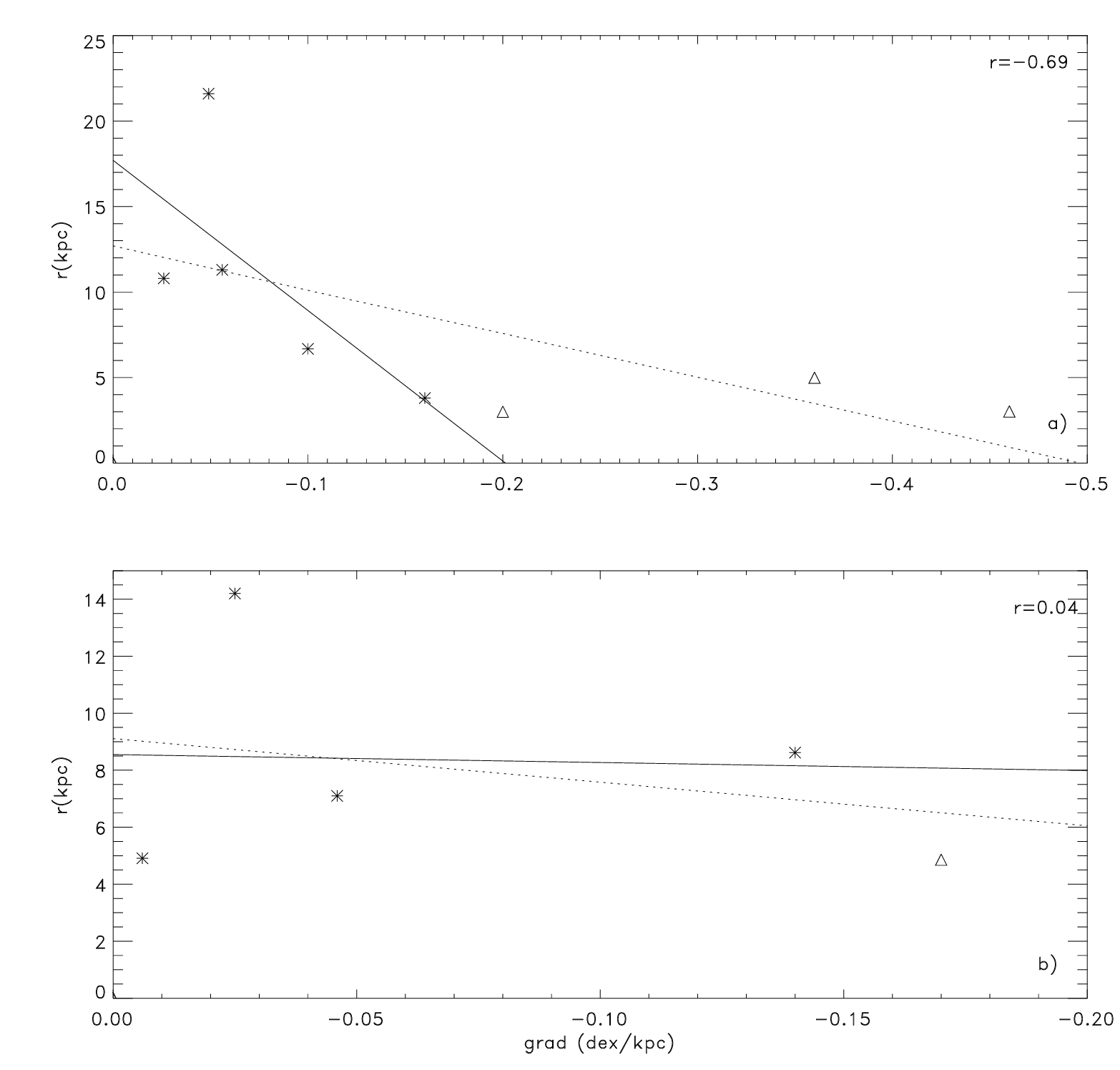}
\caption{The relationship between the optical radius and the gradient for non-barred (panel a) and barred (panel b), late-type galaxies. Symbols and lines as in figure ~\ref{fig7}.} 
\label{fig8}
 \end{figure}

\clearpage
\begin{figure}
\centering
\includegraphics[width=14cm]{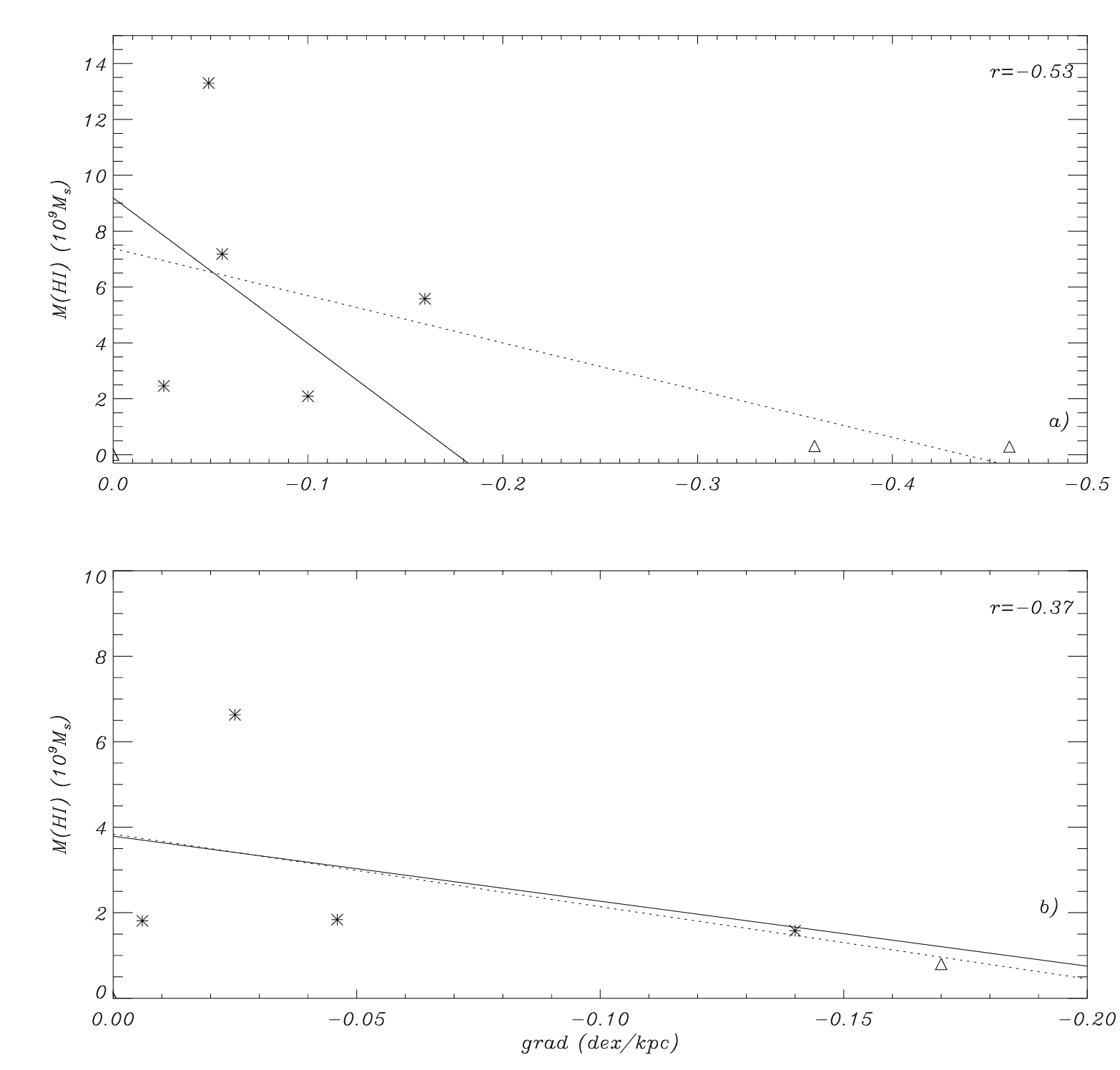}
\caption{The relationship between the gas mass and the gradients for non-barred (panel a) and barred (panel b), late-type galaxies. Symbols and lines as in figure ~\ref{fig7}.} 
\label{fig9}
 \end{figure}

\clearpage
\begin{figure}
\centering
\includegraphics[width=14cm]{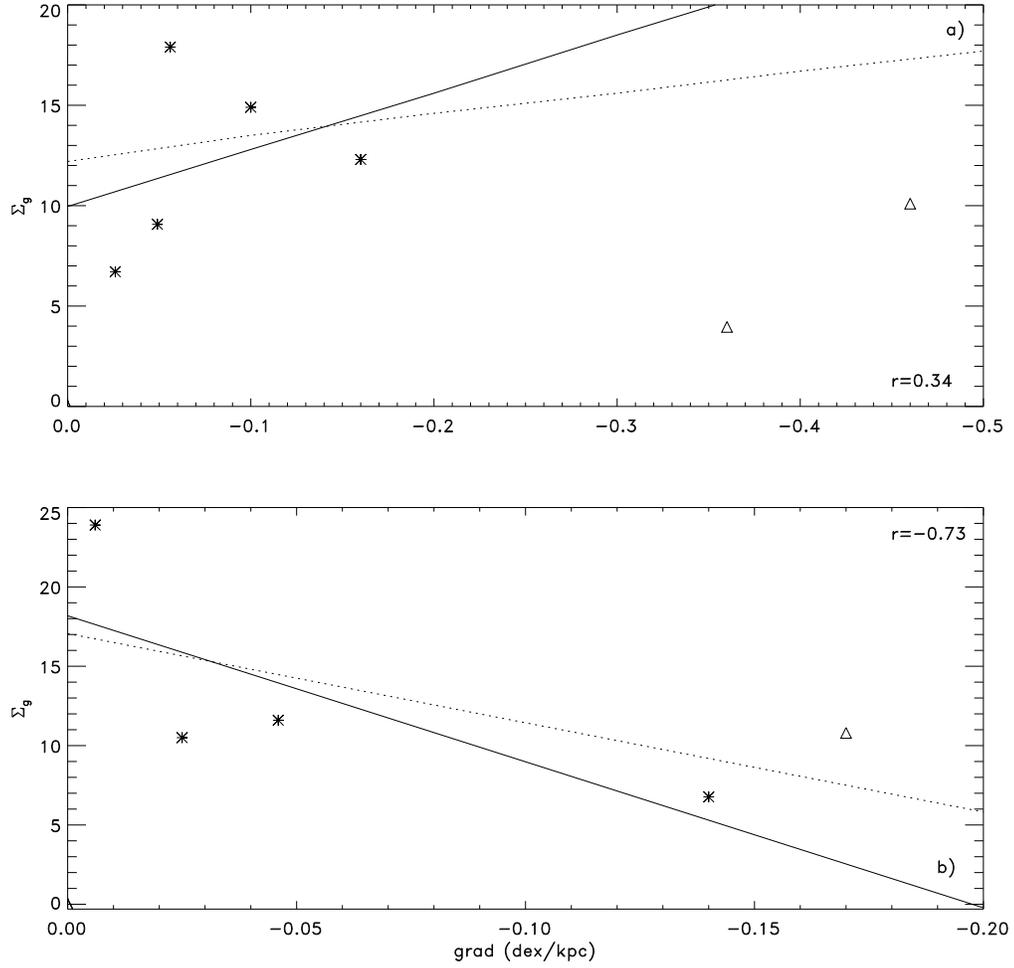}
\caption{The relationship between the gas surface density and the gradients for non-barred (panel a) and barred (panel b), late-type galaxies. $\Sigma_g$ for the dS galaxies was obtained as the gas mass divided by the area of a disc. Symbols and lines as in figure ~\ref{fig7}.} 
\label{fig10}
 \end{figure}
\end{document}